# TRANSLATION

# THERMOGRAVIMETRIC STUDY OF THE NON-STOICHIOMETRIC WÜSTITE $FeO_X$

# PSEUDO PHASES $W_i$ AND $W'_i$.  NEW  T-P-$X$  EQUILIBRIUM PHASE DIAGRAMS

# Pierre  VALLET[1], Claude  CAREL*

[1](1906-1994)    Sudoc instructions:  https://www.idref.fr/030362598#70
* **Université de Rennes 1**, 2 rue du Thabor CS 46510, 35065 RENNES Cedex, France: c-carel@orange.fr

### The present document is the result of the translation into English
### of the French publication in Ref. [∅]

**ABSTRACT –** A statistical numerical analysis of numerous thermogravimetric measurements (temperature, mass variation, oxygen pressure) obtained by P. Raccah in 1962 leads to a new equilibrium phase diagram T-P-X of the iron monoxide or wüstite written FeO$x$. The external boundaries of the stable existence field are stated with the aid of some results from the literature. Three varieties or pseudo phases $W_1$, $W_2$, $W_3$ are revealed above 911 °C, three others $W'_1$, $W'_2$, $W'_3$ below. This diagram displays their subfields. Because the transformations of one pseudo phase into another are of the second order there is no biphasic domain, except merely for the transitions $W'_j \leftrightarrows W_i$ in the vicinity of 911°C, where it exists also the first order transition α-Fe ⇆ γ-Fe. The diagram shows also 25 invariant points between stable phases and 8 possible invariant points between metastable phases.
A new thermodynamic approach since the end of the 50s is pointed out, which makes possible to characterize the point defects and their clusters as components of the crystal lattice. *ADDENDA* and Comments.

**RÉSUMÉ –** Une analyse statistique des mesures thermogravimétriques nombreuses (température, variation de masse, pression d'oxygène) obtenues par P. Raccah en 1962 mène à la description d'un nouveau diagramme de phases T-P-X à l'équilibre du monoxyde de fer ou wüstite formulée FeO$x$. Les frontières externes du domaine d'existence stable sont établies précisément avec l'aide de certains résultats de la bibliographie. Trois variétés ou pseudo-phases $W_1$, $W_2$, $W_3$ sont mises en évidence au-dessus de 911 °C, trois autres $W'_1$, $W'_2$, $W'_3$ le sont en-dessous de 911 °C. Le diagramme d'état figure leurs sous-domaines respectifs. Du fait que la transformation d'une variété en une autre est du deuxième ordre, il n'y a pas de domaine biphasé, excepté seulement pour les transitions $W'_j \leftrightarrows W_i$ au voisinage de 911 °C, où existe la transition du premier ordre α-Fe ⇆ γ-Fe. Le diagramme comprend 25 points invariants entre phases stables ainsi que 8 points invariants éventuels entre phases métastables.
Une approche thermodynamique nouvelle depuis la fin des années 50 est indiquée qui permet de caractériser les défauts ponctuels et leurs amas comme des constituants du réseau cristallin. *ADDENDA* et Commentaires.
**MOTS-CLÉS :** thermogravimétrie, hautes températures, équilibre thermodynamique, transitions du deuxième ordre, points triples, diagramme de phases

## INDEX

# I. – INTRODUCTION

This work follows the publication (1) in 1965 based on thermodynamic exploitation of Raccah's thermogravimetric results available as soon as 1962 (2) (3). It will retain the most important notations and conclusions, in particular the high temperature existence of three supposed *allotropic* varieties of non-stoichiometric iron monoxide or wüstite, referred by the symbols W$_i$: W$_1$, W$_2$, W$_3$ (4) (5). Different kinds of differentiation will confirm later these varieties called then modifications or *pseudo phases*, the denomination "allotropy" having been contested (6) (7) (8).

By dilatometry and X-ray crystallography one of us gave the first experimental proof of existence of these three pseudo phases W$_i$ (5). He showed that the passage from one pseudo phase to another is a transformation at least of second order, under defined equilibrium conditions. On the other hand, boundaries between the pseudo phases previously calculated (1) (3) (4) were confirmed (8) (9 a,b,c).

In 1966 and 1969 respectively Wagner jr *et alii* (10), and Fender and Riley (11) confirmed Raccah-Vallet-Carel's work. Authors (10) established their own experimental version of the equilibrium diagram. Following it, they verified *in situ* by conductivity the presence of varieties strictly identical to Raccah's varieties above and below 911 °C. Authors (11) explored the whole domain of stability by emf measurements. They strictly identified three varieties above 911 °C, by extrapolation only below this temperature. They confirmed their identity with [Raccah-Vallet-Carel]'s pseudo phases, and the diagram at equilibrium too.

In 1972 Hayakawa, Cohen and Reed (12) published numerous highly accurate *in situ* measurements of the cubic cell parameter as a function either of composition at several temperatures, or temperature with substantially constant composition. Analyzed from a statistical point of view at the same time as similar measurements, the existence of the three pseudo phases can be confirmed from a basic crystallographic point of view. This aspect was the subject of a short Review guided by Weigel (7) (8), and a major experimental study by neutron diffraction at temperature and composition equilibrium, with Gavarri's constant contribution (8). The short-, medium-, and long-range orders in the NaCl-type crystal disturbed by $Fe^{2+}$ vacancies and $Fe^{3+}$ ions in interstitial sites was studied also after quenching by X-ray and electron diffraction. Structurally distinct domains were proposed by Manenc (13), and Andersson and Sletnes (14) which are close to those described by us for the pseudo phases at high temperatures.

Now it always seems advisable to supplement this previous work by exploiting it more systematically with the addition of various more recent data (15 to 17). In this respect, let us cite also the compilation of Spencer and Kubaschewski (18) on the thermodynamic results published to date for the iron-oxygen system as a whole. Wüstite serves on their paper as a reference in calculations and evaluations.

## II. – WÜSTITE FORMULA AND FUNDAMENTAL PROPERTIES

Knowing that the wüstite is not stoichiometric by default of iron, it seems therefore reasonable to attribute to it a formula such as $Fe_yO$ in which $y$ is less than unity or the formula $Fe_{1-z}O$ in which $z$ is a positive number between approximately 0.05 and 0.17.

However to represent the wüstite, a formula $FeO_x$ in which $x$ is greater than the unity was used. Some authors preferred the equivalent formula $FeO_{1+u}$ in which $u$ is a positive number between 0.05 and 0.21. The composition parameter $x$ gives a much simpler mathematical representation of thermodynamic properties. Mainly the relationship between oxygen pressure $p'$ at equilibrium and composition of the wüstite $FeO_x$ is particularly simple

$$l' = \log p' = M(T)x + N(T) \qquad [1]$$

This equation established by Raccah with a good precision (2) has since been verified by many authors notably (20 to 27) with often higher precision. Of course $x$ finally varying little along an isotherm, the use of the parameter $y = 1-z$ leads to an equation of the same type as equation [1].

For all Raccah's isotherms it has been verified that the linear correlation coefficient between $l'$ and $x$ is closer to unity than the one between $l'$ and $z$ or $y$ (1) (2). The same is true if the parameter $x$ is in place of the atomic fraction $N'_O$ of oxygen in the wüstite $N'_O = x/(1+x)$.

Furthermore, it has been shown that the coefficients M(T) and N(T) in the relation [1] are linear functions of $T^{-1}$ above 911 °C for each of the three $W_i$ such as

$$M_i(T) = a_i T^{-1} + b_i \qquad [2]$$

$$N_i(T) = c_i T^{-1} + d_i \qquad [3]$$

where the index 'i' is 1, 2 or 3 depending on the pseudo phase $W_i$. It should be noted that the development with four coefficients of the function $x(l', T^{-1})$ which was used at the origin of this work (4) was taken up and applied to their thermogravimetric results on wüstite by Janowski *et alii* (27).

However, if the equation [1] is approximately verified when using $y$, $z$ or $N'_{\underline{o}}$ instead of $x$, it is not at all the same with equations [2] and [3]. The functions $M_i(T)$ and $N_i(T)$ are no longer linear in relation with $T^{-1}$. Is it any wonder that a measurement of oxygen in equilibrium with wüstite under certain conditions is better expressed as a function of the parameter of composition $x$ attributed to oxygen rather than as a function of another parameter of composition relating to iron such that $y$ or $1-z$?

To date (1979) it seems accepted that the complete interpretation of the properties of wüstite can only be obtained by structural, electronic and other research means from the solid state physics-chemistry. As a result, many researchers such as (28) have embarked on this path.

### III. – SOLID WÜSTITES. T-P-X DIAGRAMS

It has already been pointed out (29) (30) that the equilibrium phase diagram of the wüstite originally given (3) has to be modified, so it has to be retouched presently.

1.– <u>External boundaries of the solid domain</u>.

a – <u>Iron/Wüstite boundary</u>. A fine analysis of a set of measurements of $l'_{o}$ showed that it could be separated into two subsets by the isotherm at 911 °C, temperature of the $\alpha$-⇆$\gamma$-Fe transformation. This was consistent with the obvious fact from Raccah's thesis (2) that the wüstite does not get the same properties on both sides of 911 °C (30 a,b). So it will be represented by the symbol W at temperatures above and W' below 911 °C.

Results from various authors could be added to ours, and give a satisfactory description of the external boundaries. If the results of series of measurements obtained by various experimenters are mixed without precautions, only a more or less crude average description of little interest may be obtained. So a meticulous statistical sorting has to be performed.

Above 911 °C, the $\gamma$-Fe/W boundary appears to be best represented by the equation obtained from 55 Rizzo's results (17)

$$l'_{o\gamma} = -27\,460\,T^{-1} + 6.740\,1 \qquad [4]$$

providing a confidence interval $\Delta l'_{o\gamma} = \pm 0.012$. At 911°C, this equation gives $l'_{o\gamma} = -16.452$.

Below 911 °C, the $\alpha$-Fe/W' boundary is represented by the equation already given (30 a)

$$l'_{o\alpha} = -27\,856\,T^{-1} + 7.027\,5 \qquad [5]$$

providing $l'_{o\alpha}$ with a confidence interval $\Delta l'_{o\alpha} = \pm 0.020$. At 911 °C it provides $l'_{o\alpha} = -16.500$.

b − <u>Wüstite/Fe3O4(Magnetite) boundary</u>. It has been shown that this boundary can be represented by a fairly complicated four terms equation. Its analytical determination has been facilitated by new measurements (15) (9-c)) following

$$l'_1 = -\,56\,575.05\,T^{-1} - 39.608\,201\,\ln T\, +\, 0.016\,712\,775\,T\, +\, 293.591\,065 \qquad [6]$$

This equation was calculated using 22 measurements, 4 of which being due to Darken and Gurry (31) and the others to Raccah (2) and Carel (9 c). The representative curve has an inflection point for $T_{infl} = 1\,185$ K, or 912 °C. Equation [6] provides $l'_1$ such as $\Delta l'_1 = \pm\,0.021$.

The determination of energetic properties (work functions) (32) between 675 and 950 °C provided values of $l'$(T) on the external boundaries in the diagram. They could be exploited.

2.− <u>Equilibrium phase diagram in coordinates $(T^{-1}, l')$</u>

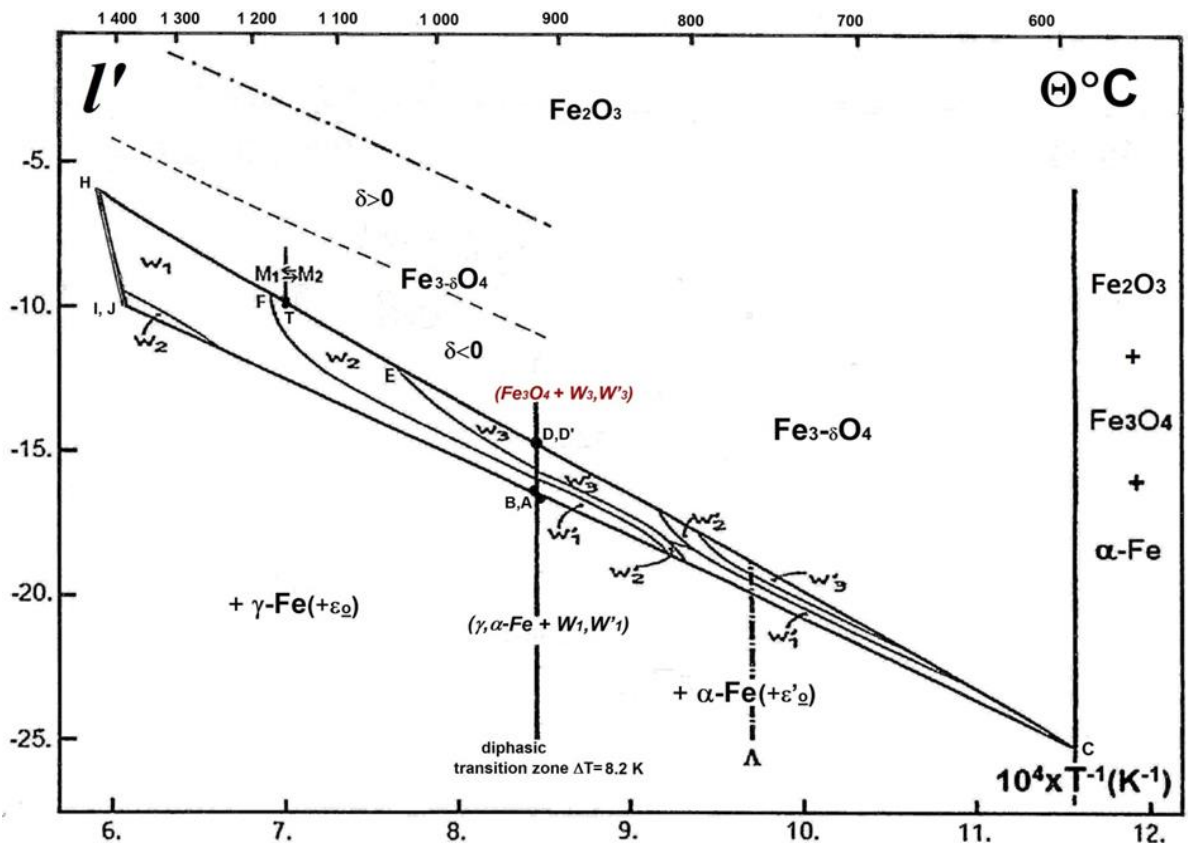

Fig. 1- Equilibrium solid wüstite diagram in coordinates $(T^{-1}, l')$

The vertical line at the triple points B and A is at the place of the transitions α-Fe⇆γ-Fe, and W⇆W'; is it an equivalent (Fe3O4 + W3,W'3) at D,D'? The triple point T at 1 433 K is formed of (W2 + the first order transition M1⇆M2 of Fe3O4); the Λ transition of α-Fe is continuously present across the W' domain; the boundary Fe3O4/Fe2O3 and the dashed line for δ=0 are from R. Dieckmann; See **V.− ADDENDA** §1 to 3 a -d.

The previous results lead to the correction of the preceding diagrams [See *ADDENDA*, §1 p. 17 and (29) Fig. 2 p. 248]. Figure 1 above shows the new version of the diagram $(T^{-1}, l')$. At 911 °C, it has three invariant points A, B and D. Their respective ordinates are $l'_A = -16.452$, $l'_B = -16.500$ and $l'_{D,D'} = -14.698$.

At the triple point B, $\alpha$-Fe, $\gamma$-Fe and W' coexist in equilibrium. The same applies at the triple point A for $\gamma$-Fe, W and W', or possibly at the point D-D' for $Fe_3O_4$, W and W'.

The arc BC of the boundary W'/$\alpha$-Fe defined by equation [5], and the arc DC of the boundary W'/ $Fe_3O_4$ defined by equation [6] intersect at the invariant Chaudron's point C (33) whose temperature is obtained by the following equation

$$l'_{o\alpha} \ = \ l'_1 \qquad\qquad [7]$$

from equations [5] and [6].

Equation [7] gives $T_C$ = 869.4 K or 596.2 °C, which would be consistent with [Bonneté, Païdassi]'s experiments (34) locating this temperature between 580 and 600° C (See *ADDENDA 3–d*)).

The ordinate of point C corresponding to equation [7] is $l'_C = -25.013$. It gives an oxygen pressure $p'_C$ of about 9.7 $10^{-26}$ atm, a value so low that at 869.4 K it would take a volume of 1 500 liters of the mixture carbon monoxide/carbon dioxide under one atmosphere to meet on average an isolated molecule of oxygen. This means that the oxygen pressure no longer makes any sense. The obvious significance of the regions in the diagram is recalled in Figure 1.

3.– Equilibrium diagram in coordinates $(x,\ \Theta°C)$

a– Underline General considerations. It has already been shown (1) that under equations [1], [2] and [3] the decimal logarithm $l'$ of the oxygen pressure $p'$ in equilibrium at the absolute temperature T with a solid wüstite $W_i$ formulated $FeO_x$ is given above 911 °C by the equation

$$l' \ = \ \log p' = (a_i T^{-1} + \ b_i)\, x + c_i T^{-1} + d_i \qquad\qquad [8]$$

On the other hand, since the equilibrium pressure $p'$ of oxygen is always less than $10^{-6}$ atm, this oxygen can be assimilated to a perfect gas, so that it also obeys the following equation

$$l' \ = \ \triangle \bar{G}'/(19.144\ 64\ T) \qquad\qquad [9]$$

in which $\Delta \bar{G}'$ represents the partial molar free energy expressed in joule of the dissolution of oxygen in $W_i$. In these two equations, the reference state of oxygen is that of molar gaseous oxygen under one atmosphere. In equation [9] and later, the value R ln 10 = 19.144 64 J.K$^{-1}$.mole$^{-1}$ was used according to Rossini (35).

When multiplying the two members of equation [8] by T, the function obtained is

$$Z_i = (a_i + b_i T)\, x + c_i + d_i\, T \qquad\qquad [10]$$

equivalent to the function

$$Z_i = l'T = \Delta \bar{G}' / 19.144\ 64 \qquad\qquad [11]$$

proportional to the chemical potential of oxygen in $W_i$. $Z_i$ is thus a function of the second degree with respect to $x$ and T. It is therefore represented by a quadric in the space with three dimensions $x$, T and $Z_i$. It has been shown that this quadric is an hyperbolic paraboloid whose rectilinear generators are the isotherms and the lines of equal composition (9 a).

b– Determination of the  coefficients in equation [8]

1st method. It was used only for the three $W_i$. It makes direct use of equations [2] and [3] when the coefficients $M_i$ and $N_i$ have been calculated for several isotherms of a wüstite $W_i$. This direct method provided the first assessment (1) of the coefficients in equation [8].

2nd method. It was applied first (30 a) to the three $W'_i$ for which the first method is unusable, then to the three $W_i$ (30 b,c).

In the case of the three $W'_i$, $l'$ is calculated using Raccah's isotherms for various values of $x$ increasing in increments of 0.0025 between 1.055 and 1.115 at the 11 experimental temperatures (1). It is found that the relation

$$l' = A(x)\, T^{-1} + B(x) \qquad [12]$$

is well followed. Thus the coefficients are distributed into three groups defining three $W'_i$. In first approximation for each of them $A(x)$ and $B(x)$ are linear functions noted

$$A_i = a_i x + c_i \qquad [13]$$

$$B_i = b_i x + d_i \qquad [14]$$

It results that

$$l' = (a_i x + c_i)\, T^{-1} + b_i x + d_i \qquad [15]$$

is an equation identical to equation [8]. The same mathematical treatment can therefore be applied to the three $W_i$.

3rd method. It differs from the previous one only by the prior selection of the isotherms to which it is applied (See *ADDENDA* §1 – Metastability and sorting out the isotherms, Figs. a) and b)).

Case of the 3 $W_i$. The isotherms relating to a given wüstite $W_i$ are selected in three groups by the 1st method. The second method is then applied to each of them. It is found that equations [13] and [14] define linear correlations between $A_i$ or $B_i$ and $x$ with coefficients very close to the unit (all greater than 0.999 999 99) so that in Table 1, $a_i$ and $c_i$, are defined with a confidence interval of about $\pm\, 0.3$, and $b_i$ and $d_i$ with a confidence interval of about $\pm\, 0.000\ 3$.

Case of the 3 $W'_i$.  In reality the approximation defined by equations [13] and [14] can be improved by replacing them with the following second degree equations

$$A_i = a'_i x^2 + c'_i x + e'_i \qquad [16]$$

$$B_i = b'_i x^2 + d'_i x + f'_i \qquad [17]$$

The result is the equation

$$l' = (a'_i x^2 + c'_i x + e'_i)\, T^{-1} + b'_i x^2 + d'_i x + f'_i \qquad [18]$$

On the other hand, the selection of the segments of the isotherms on which the calculation by the 2nd method is into focus plays an essential role (See *ADDENDA* §2-Sorting out). For $W'_1$ these are segments of the isotherms at 880, 875, 865, 820 and 800 °C, for $W'_2$ segments of the isotherms at 900, 880, 870, 855, 840, 820 and 800 °C, and for $W'_3$ segments of the isotherms at 900, 875, 870, 865, 855, 840, 820 and 800 °C. Of course, in the calculation of $l'$ when using equation [1] for $x$ placed at regular values, appropriate values of $M_i$ and $N_i$ are needed. They are those of segments of the isotherms close to the $\alpha$-Fe/W' boundaey for $W'_1$ or $W'_2$ except at 880°C for the latter, and those of segments near the $W'/Fe_3O_4$ boundary for $W'_2$ at 880 °C or for $W'_3$.

Finally at 820°C the isotherm is better represented by the following equation than by two successive segments of straight lines

$$l' = 72.805\,294\,x^2 - 136.569\,99\,x\ +\ 44.645\,454 \qquad [19]$$

The same applies to 800 °C where the isotherm is represented by the equation

$$l' = 105.528\,70\,x^2 - 207.770\,87\,x\ +\ 82.859\,59 \qquad [20]$$

The confidence interval of $l'$ is $\pm 0.027\,2$ for equation [19] and $\pm 0.022\,5$ for equation [20].

As in the case of the three $W_i$ the coefficients of equation [18] are obtained with great precision. This is why it is possible to give them with six perfectly justified significant figures.

<u>Note</u>. The high accuracy of the results of this 3rd method shows that the isotherms define accurately hyperbolic paraboloids for each of the three $W_i$, and 3rd degree ruled surfaces for the three $W'_i$.

<u>Table 1</u>: coefficients of equation [8]          c − <u>Results of calculations</u>

| | $a_i$ | $b_i$ | $c_i$ | $d_i$ |
|---|---|---|---|---|
| $W_1$ | 46 753.4 | - 7.378 1 | - 78 825.3 | 16.061 3 |
| $W_2$ | - 9 568.9 | 31.172 8 | - 18 413.3 | - 25.256 9 |
| $W_3$ | - 33 238.9 | 48.366 9 | 6 883.9 | - 43.566 9 |
| coefficients by the 2$^{nd}$ method | | | | |
| $W'_1$ | - 10 260 | 27.960 | - 18 155.1 | - 21.386 6 |
| $W'_2$ | - 2 700 | 23.700 | - 26 189.5 | - 16.895 5 |
| $W'_3$ | - 21 580 | 44.240 | - 5 607.7 | - 39.294 6 |

Table 1 gives the coefficients of equation [8] as obtained by the 3$^{rd}$ method for the $W_i$ (30 b) and the 2$^{nd}$ method for the $W'_i$ (30 a).

Table 2 gives the coefficients of equation [18] calculated for the $W'_i$ by the 3$^{rd}$ method.

d − <u>Determination of the external boundaries of the diagram</u>

The principle of this determination is simple and it is the same regardless of the pseudo phases $W_i$ or $W'_i$.

It should be noted that the accurate rectification of these boundaries is essential when computing the molar thermodynamic properties $H_i$, $S_i$, $C_{pi}$ from the thermogravimetric results (See *ADDENDA* § 3 −d).

| | $W'_1$ | $W'_2$ | $W'_3$ |
|---|---|---|---|
| $a'_i$ | 1 718 126.728 | 1 372 284.000 | 1 419 479.827 |
| $b'_i$ | - 1 498.280 093 | - 1 195.160 000 | - 1 239.113 939 |
| $c'_i$ | - 3 655 355.221 | - 2 987 723.828 | - 3 156 098.859 |
| $d'_i$ | 3 205.158 689 | 2 623.569 380 | 2 780.570 131 |
| $e'_i$ | 1 914 755.682 | 1 597 150.861 | 1 724 745.998 |
| $f'_i$ | - 1 705.377 467 | - 1 430.856 179 | - 1 549.890 140 |

In the case of the boundary with $Fe_3O_4$ at any selected temperature T, the value of $l'_1$ given by equation [6] is identified with the value of $l'$ relating to a given wüstite $W_i$ or $W'_i$, provided either by equation [8] or equation [18] with its own coefficients, either from Table 1 or Table 2.

In the case of the boundary with $\gamma$-Fe, the value of $l'_{o\gamma}$ given by equation [4] is identified with the value of $l'$ at the same temperature and relating to a given variety $W_i$, provided by equation [8] with its own coefficients extracted from Table 1.

For the boundary with $\alpha$-Fe, the value of $l'_{o\alpha}$ given by equation [5] is identified with the value at the same temperature and relating to a given variety $W'_i$ defined by equation [8] or [18], with its own coefficients extracted from either Table 1 or Table 2.

e – Determination of the internal boundaries between varieties taken two by two

Between two varieties $W_i$ and $W_j$ or $W'_i$ and $W'_j$ there is a boundary whose general equation is

$$l'_i = l'_j \qquad [21]$$

In the case of $W_i$ and $W_j$, $l'_i$ and $l'_j$ are given by equation [8] provided by Table 1 with the appropriate coefficients. The equations of type [21] have already been given for the three boundaries $W_1/W_2$, $W_2/W_3$ and $W_3/W_1$ (30-b)).

In the case of $W'_i$ and $W'_j$, if $l'_i$ and $l'_j$ are given by equation [8] the equation of type [21] will draw its coefficients from Table 1 as in the previous case, and the equations of the three boundaries $W'_1/W'_2$, $W'_2/W'_3$ and $W'_3/W'_1$ have already been given. If instead, equation [18] and the coefficients in Table 2 are used, for each temperature an equation of the second degree in $x$ for each internal boundary is obtained.

Given the temperature $\Theta$ expressed in degree Celsius, Figure 2 provides a general view of the phase diagram. Obtained by the 3rd method above 911 °C, it is very similar to the one previously presented (3). The isotherm at 911 °C not only shares $\alpha$-Fe and $\gamma$-Fe, but the whole diagram, since on both sides of this isotherm the properties of wüstite are different (15).

Fig. 2 – Equilibrium diagram of solid wüstite in coordinates ($x$, $\Theta$°C).
Crosses represent the breaks observed on the isotherms obtained by Raccah (1) (2).
The dashed line from point T corresponds to a first order transition $M_2 \leftrightarrows M_1$ of Fe3O4 at 1 160°C.
Transformation $\Lambda$ in $\alpha$-Fe meets again across the domain of W' (See *ADDENDA* § 3 –d)).

The hyperbolas defined by the equations previously given (3) (9) (30 b) represent the boundaries $W_1/W_2$, $W_2/W_3$ and $W_3/W_1$. This latter being metastable is represented in dashes. These three hyperbolas consist of two arcs symmetrical regarding their respective centers. The experimental confirmation of the existence of higher arcs at high temperatures and low values of $x$, and that of lower arcs at lower temperatures and higher values of $x$ was obtained by Xray-crystallography with the collaboration of Weigel (36). The confirmation of lower arcs was obtained mainly by dilatometry by Carel, in part with the collaboration of Bars (5) (9 a,b,c).

The first confirmation of the pseudo phases outside Vallet's group was obtained in 1966 with *in situ* measurements of electrical conductivity by Wagner *et alii* (10). The second remarkable confirmation of the pseudo phases including the boundaries between them was brought by Fender and Riley in 1969, by means of emf measurements in solid electrolyte batteries (11) (See ENDNOTES 1- p. 19).

The $\gamma$-Fe/$W_i$ boundary consists of three arcs AG, GI and IJ along which the $\gamma$-Fe is in equilibrium successively with the wüstites $W_1$, $W_2$ then $W_3$. The last arc IJ is very short because the existence domain of $W_3$ at high temperature is extremely narrow. It is reduced to a point in Figure 2. Similarly the boundary $W_i$ /Fe₃O₄ consists of three consecutive arcs DE, EF and FH along which Fe₃O₄ is in equilibrium successively with $W_3$, $W_2$ and $W_1$. The order of indices is reversed from what it is on the $\gamma$-Fe/$W_i$ boundary at increasing temperature.

Darken and Gurry (31) determined the temperature 1 371 °C or 1 644 K of invariant point J where $\gamma$-Fe, liquid wüstite $W_L$ and solid $W_3$ defined by $x$ = 1.045 05 are in equilibrium under an oxygen pressure defined by $l'$ = -9.973. The same authors give the temperature 1 424 °C or 1 697 K of the invariant point H where Fe₃O₄ is in equilibrium with the liquid wüstite $W_L$ and the solid $W_1$ defined by $x$ = 1.212 14 under an oxygen pressure (expressed in atmosphere) such as $l'$ = -5.937.

<u>Table 3</u> — Invariant points in diagram T-p($O_2$)-$x$ of W

| Point | Phases | T (K) | Θ (°C) | $x$ | $l'$= $\log_{10}p$ ' |
|---|---|---|---|---|---|
| | | 1) Equilibria between stable phases | | | |
| A | $W_1$, $W'_1$, $\gamma$-Fe | 1 184 | 911 | 1.060 8 | -16.452 |
| Q | $W_1$, $W'_1$, $W_2$ | 1 184 | 911 | 1.076 2 | -15.959 |
| R | $W_2$, $W'_3$, $W_3$ | 1 184 | 911 | 1.092 4 | -15.585 |
| D | $W_3$, Fe₃O₄ | 1 184 | 911 | 1.136 1 | -14.698 |
| D' | $W'_3$, Fe₃O₄ | | | 1.127 6 | |
| E | $W_2$, $W_3$, Fe₃O₄ | 1 310 | 1 037 | 1.144 3 | -12.001 |
| F | $W_1$, $W_2$, Fe₃O₄ | 1 447 | 1 174 | 1.158 4 | -9.569 |
| H | $W_1$, $W_L$, Fe₃O₄ | 1 697 | 1 424 | 1.212 1 | -5.937 |
| G | $W_1$, $W_2$, $\gamma$-Fe | 1 510 | 1 237 | 1.047 1 | -10.717 |
| I | $W_2$, $W_3$, $\gamma$-Fe | 1 643 | 1 370 | 1.045 1 | -9.969 |
| J | $W_3$, $W_L$, $\gamma$-Fe | 1 644 | 1 371 | 1.045 0 | -9.973 |
| K | $W_2$, $W_3$, $W_L$ | 1 644 | 1 371 | 1.045 1 | -9.966 |
| L | $W_1$, $W_2$, $W_L$ | 1 653 | 1 380 | 1.065 4 | -9.503 |
| | | 2) Possible metastable equilibria | | | |
| 1 | $W_1$, $W_3$, $\gamma$-Fe | 1 548 | 1 275 | 1.046 8 | -10.967 |
| 2 | $W_1$, $W_3$, $W_L$ | 1 648 | 1 375 | 1.057 5 | -9.660 |
| 3 | $W_1$, $W_3$, Fe₃O₄ | 1 403 | 1 130 | 1.152 2 | -10.332 |
| 4 | $W_1$, $W_3$, $W'_2$ | 1 184 | 911 | 1.080 0 | -15.836 |

On the other hand, since the general diagram of iron oxides published by Darken and Gurry (31) serves as a model, it can be admitted that the boundary between solid wüstite $W_i$ and liquid wüstite $W_L$ is a straight line.

It can be represented in Figure 1 by line JH having for equation

$$l' = -2.124\ 5\ x \times 10^5\ T^{-1} + 119.255 \qquad [22]$$

The same boundary in Figure 2 has for equation

$$T = 317.19\ x + 1\ 312.5 \qquad [23]$$

The other invariant points in the diagram have an obvious meaning. It is sufficient to collect them in Table 3 by indicating for each of them the nature of the condensed phases in equilibrium with the gas phase as well as the temperature, the decimal logarithm of oxygen pressure expressed in atmosphere, and the composition of the solid $W_i$ ensuring this equilibrium.

Below 911 °C with the representation mode of $l'$ selected at first time (2nd method), the diagram appears not simple in Fig. 3 especially in the vicinity of point C.

The boundary between $\alpha$-Fe and W' is formed of two consecutive arcs BG' along which $\alpha$-Fe is in equilibrium with $W'_1$, and G'C along which $\alpha$-Fe is in equilibrium with $W'3$. The boundary between $Fe_3O_4$ and W' is made up of three consecutive arcs CF', F'E' and E'D'. Along the two arcs CF' and E'D' $Fe_3O_4$ is in equilibrium with $W'_3$, and along the arc F'E' $Fe_3O_4$ is in equilibrium with $W'_2$.

The internal boundary $W'_1$/ $W'_2$ is represented by the arc Q'S. The internal boundary $W'_2$/ $W'_3$ is represented by the two arcs R'E' and SF'. Finally, the internal boundary $W'_1$/ $W'_3$ is represented by the arc G'S, and its metastable extension the dotted line SS'. The invariant point S is the junction point of the three arcs G'S, F'S and Q'S where coexist the three wüstites $W'_1$, $W'_2$ and $W'_3$.

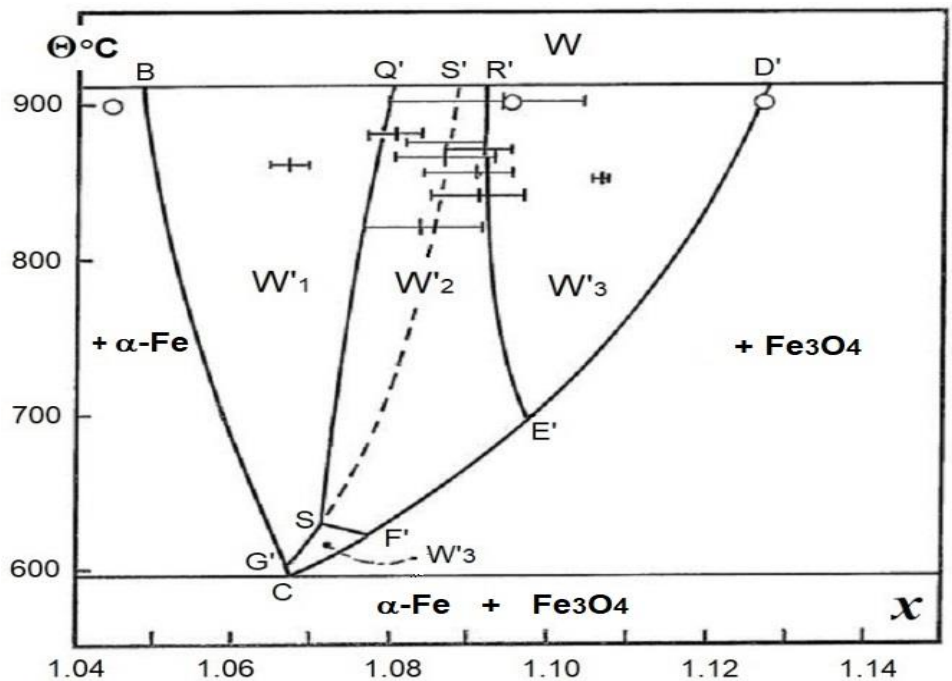

Fig. 3 − Equilibrium diagram of W' (2$^{nd}$ method). Horizontal segments are the statistical confidence limits of the angular points on Raccah's isotherms.

The diagram is divided into four regions. The domain of stable existence of $W'_1$ is the area BG'SQ', that of $W'_2$ the area Q'SF'E'R', and that of $W'_3$ is made of two areas R'E'D' and CG'SF'. At Chaudron's point C, $W'_3$ is in equilibrium with $\alpha$-Fe and $Fe_3O_4$.

Table 4 below gives the coordinates of the main invariant points of W'. It is analogous to Table 3. There is only one possible metastable equilibrium invariant point S' in it, while the diagram obviously contains several other such points, but it is the sole to be interesting.

<u>Table 4</u> – Invariants points in the T-$p'$-$x$ diagram of W' (2nd method)

| Point | Phases | T (K) | $\Theta$ °C | $x$ | $l' = \log_{10} p'$ |
|---|---|---|---|---|---|
| | | 1) Equilibria between stable phases | | | |
| B | W'$_1$, $\alpha$-Fe, $\gamma$-Fe | 1 184 | 911 | 1.048 0 | -16.500 |
| Q' | W$_2$, W'$_1$, W'$_2$, | 1 184 | 911 | 1.079 8 | -15.886 |
| R' | W$_2$, W'$_2$, W'$_3$, | 1 184 | 911 | 1.091 8 | -15.629 |
| D' | W$_3$, W'$_3$, Fe3O4 | 1 184 | 911 | 1.127 6 | -14.698 |
| G' | W'$_1$, W'$_3$, $\alpha$-Fe | 875 | 602 | 1.067 4 | -24.808 |
| S | W'$_1$, W'$_2$, W'$_3$, | 902 | 629 | 1.071 5 | -23.743 |
| F' | W'$_2$, W'$_3$, Fe3O4 | 893 | 620 | 1.077 2 | -23.960 |
| E' | W'$_2$, W'$_3$, Fe3O4 | 968 | 695 | 1.097 5 | -23.000 |
| C | W'$_3$, $\alpha$-Fe, Fe3O4 | 869.4 | 596.2 | 1.067 6 | -25.013 |
| | | 2) Possible metastable equilibrium | | | |
| S' | W$_2$, W'$_1$, W'$_3$, | 1 184 | 911 | 1.088 0 | -15.728 |

The diagram in Figure 2 appears not to be appropriate to represent the phenomena produced at 1 184 K or 911 °C. Considering the plane perpendicular to the plane $x$OT, its trace is the line BD in Figure 2. This isothermal plane corresponds to T = 1 184 K. Figure 4 shows the variations of $l'$ as a function of $x$ in this plane. Each wüstite W$_i$ or W'$_i$ gives a linear segment. Then the isotherm of W consists of three segments AQ, QR and RD. Similarly the isotherm of W' consists of three segments BQ', Q'R' and R'D'. For a given value of $l'$ two values of $x$, $x_i$ for W$_i$ and $x'_i$ for W'$_i$, are obtained. There are two points of which W and W' have the same composition for two particular values of $l'$, these are M$_1$ defined by $x_1$=1.076 4 and $l'_1$= -15.952, and M$_2$ defined by $x_2$= 1.097 5 and $l'_2$= -15.480. These points M$_1$ and M$_2$ are not reported in Figs. 3 and 5 to avoid overloading them.

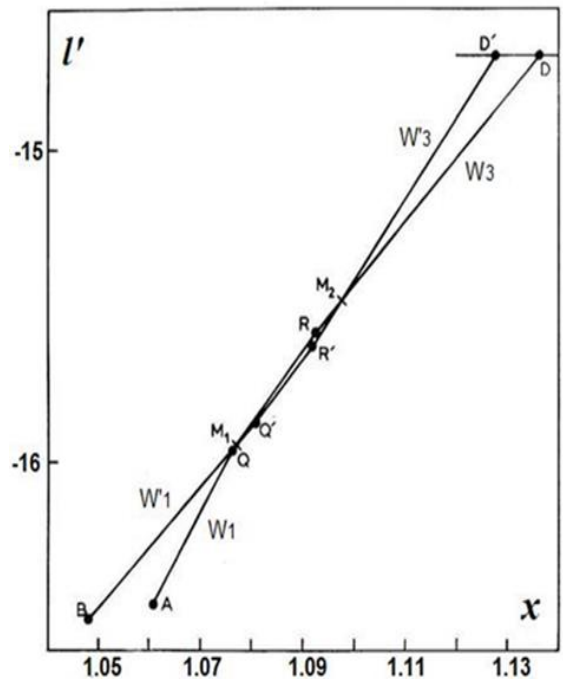

Fig. 4 – Equilibrium diagram reduced to the isotherm at 1 184 K in coordinates ($l'$, $x$) (See Vallet (30); *ADDENDA* § 3–a, b)

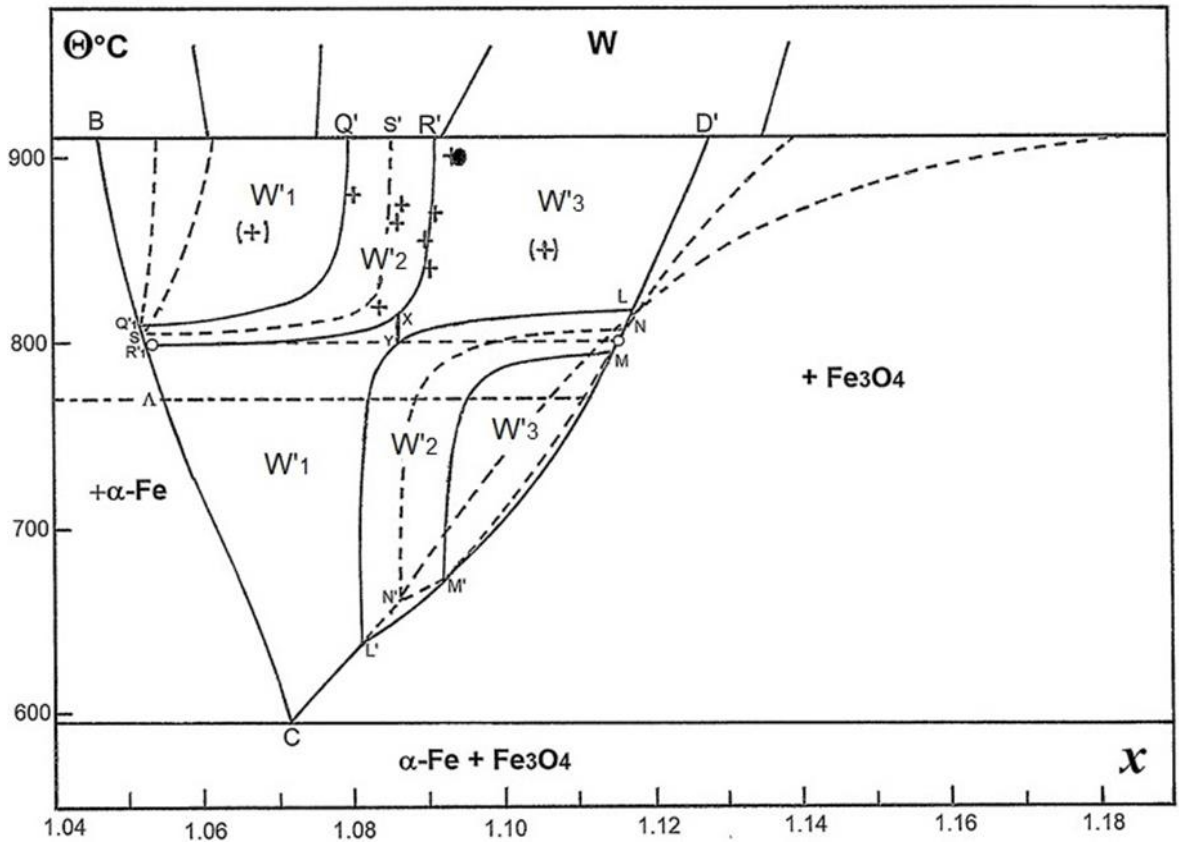

Fig. 5 −The W' phase diagram in coordinates ($x$, $\Theta$ °C) according to the 3$^{rd}$ method.
The symbols '+'are for points of separation of the two parts of the isotherms.
Symbols ➥ and ○ are for points from Dodé *et alii* 's results.

This figure gives a new diagram of W', based on the use of the 3$^{rd}$ method and the equations [16] [17] and [18]. Its internal and external boundaries are determined in the same manner as in the previous diagrams. The internal boundaries are no longer equilateral hyperbolas but more complicated curves of which the general equation derived from equation [21] is as following

$$T = (A_i − A_j) / (B_j − B_i)$$ [24]

or else 
$$T = \frac{(a'_i − a'_j)\, x^2 + (c'_i − c'_j)x\ + e'_i − e'_j}{(b'_j − b'_i)x^2 + (d'_j − d'_i)x + f'_j − f'_i}$$ [25]

whose coefficients are given in Table 2.

The curves representing the functions T($x$) of type [25] are varied. Fortunately in the diagram in Figure 5, $x$ is varying in a rather small interval. The three internal boundaries are formed each of two arcs of hyperbolic curve assuming a vertical asymptote. Table 5 gives the coordinates of the most important points in this diagram with the same indications as in Tables 3 and 4.

<u>Table 5</u> – Invariant points in T-p(O₂)-$x$ diagram of W' (3$^{rd}$ method) in Fig. 5

| Point | phases | T (K) | $\Theta° C$ | $x$ | $l'$ |
|---|---|---|---|---|---|
| | 1) Equilibria between stable phases | | | | |
| B | $\alpha$-Fe, $\gamma$-Fe, W'$_1$ | 1 184 | 911 | 1.045 8 | -16.429 |
| Q' | W$_2$, W'$_1$, W'$_2$ | 1 184 | 911 | 1.079 5 | -15.903 |
| R' | W$_2$, W'$_2$, W'$_3$ | 1 184 | 911 | 1.091 2 | -15.651 |
| D' | W$_3$, W'$_3$, Fe3O4 | 1 184 | 911 | 1.128 4 | -14.698 |
| Q'$_1$ | $\alpha$-Fe, W'$_1$, W'$_2$ | 1 082 | 809 | 1.051 5 | -18.719 |
| R'$_1$ | $\alpha$-Fe, W'$_2$, W'$_1$ | 1 073 | 800 | 1.052 2 | -18.919 |
| L | W'$_2$, W'$_3$, Fe3O4 | 1 091 | 818 | 1.117 5 | -17.134 |
| X | W'$_3$, W'$_2$, W'$_1$ | 1 086 | 813 | 1.085 7 | -17.984 |
| Y | W'$_1$, W'$_2$, W'$_3$ | 1 068 | 795 | 1.085 7 | -18.424 |
| M | W'$_2$, W'$_3$, Fe3O4 | 1 066 | 793 | 1.115 2 | -17.787 |
| M' | W'$_2$, W'$_3$, Fe3O4 | 949 | 676 | 1.092 9 | -21.688 |
| L' | W'$_1$, W'$_2$, Fe3O4 | 923 | 650 | 1.081 1 | -22.775 |
| **\*C** | **W'$_1$, $\alpha$-Fe, Fe3O4** | **869.4** | **596.2** | **1.071 3** | **-25.013** |
| | 2) Possible metastable Equilibria | | | | |
| S' | W$_2$, W'$_1$, W'$_3$ | 1 184 | 911 | 1.085 2 | -15.813 |
| S | $\alpha$-Fe, W'$_1$, W'$_3$ | 1 077 | 804 | 1.052 0 | -18.832 |
| N | W'$_1$, W'$_3$, Fe3O4 | 1 083 | 810 | 1.117 7 | -17.294 |
| N' | W'$_1$, W'$_3$, Fe3O4 | 937 | 664 | 1.086 0 | -22.170 |

*For a better adjustment of the boundary $\alpha$-Fe/W'$_1$ particularly the Chaudron's point, see *ADDENDA* § 3 –d).

The interpretation of this new diagram is more delicate than that of the previous ones. In particular, the domains of existence of W'$_3$ above 800 °C and W'$_1$ below appeared to be open to each other, without separation boundary. On the other hand, the arc LL' of the boundary W'$_1$/W'$_2$ appeared above 800 °C to separate W'$_3$ from W'$_2$ and not W'$_1$ from W'$_2$.

<u>Table 6</u> – $l'$ calculated at 800 °C for the three W'$_i$ and different values of $x$

| $x$ | 1.075 | 1.085 | 1.0875 | 1.095 | 1.110 |
|---|---|---|---|---|---|
| W'$_1$ | -18.533 | -18.325 | -18.269 | -18.095 | -17.713 |
| W'$_2$ | -18.523 | -18.322 | -18.270 | -18.105 | -17.748 |
| W'$_3$ | -18.519 | -18.318 | -18.264 | -18.099 | -17.740 |

<u>Table 7</u> – $l'$ calculated on arc LY of the boundary W'$_1$/W'$_2$, and for W'$_3$

| T (K) | 1 078 | 1 083 | 1 085 | 1 087 | 1 089 |
|---|---|---|---|---|---|
| $x$ | 1. 090 2 | 1. 096 1 | 1.100 0 | 1. 105 0 | 1. 111 5 |
| arc LY | -18.086 | -17.831 | -17.693 | -17.527 | -17.322 |
| W'$_3$ | -18.071 | -17.812 | -17.671 | -17.499 | -17.286 |

This anomaly was later resolved by defining the segment XY on the vertical asymptote of the metastable boundary W'$_1$/W'$_3$ (See *ADDENDA*, § 3-c)). This is why the points X and Y, and the segment XY appears now in Table 5 and Figure 5.

Like many spatial diagrams projected on a plane, that in figure 5 suffers from not having the coordinate *l'* perpendicular to the plane of the figure. Nevertheless the calculation of *l'* for different values of *x* at 800 °C and for each of the three W'$_i$ leads to Table 6. It should then be remembered that *l'* can be known experimentally except for a few units (approximately three) of the second decimal order so that it is practically impossible to distinguish the three W'$_i$ one from another at 800 °C. For the arc LY of LL' it is more or less similar. Table 7 gives *l'* along the horizontal part of the boundary W'$_1$/W'$_2$ (arc LY) at different values of *x* and the corresponding value for W'$_3$. It is found that up to *x* ∼ 1.105 it is difficult to distinguish the three W'$_i$. This singular property is represented in Figure 5 by a dashed line. Below 800 °C, the differences between the three W'$_i$ are rapidly increasing.

The breaks on Raccah's isotherms are marked on Figures 2, 3 and 5 by crosses. They are placed very close to the internal boundaries except at the temperatures of 850 and 860°C where they remain unexplained in all the sorting schemes (See above "3$^{rd}$ method, correspondence Θ °C ⇔ W'$_i$ "; ADDENDA § 1 −sorting the isotherms for W').

In Figure 3 the confidence intervals +Δ*x* and -Δ*x* on the determination of abscissa *x* are added to these crosses. The calculation was made according to the method already described (See Ref. (1) p. 27, § 2−"intervalle de confiance") using the linear representation of isothermal portions according to equation [1].

The Chaudron's point C in Figure 5 corresponds to *x* = 1.071 3 or *y* = 0.973 4 of which the complement to unity *z* = 0.066 55 is close to 1/15, a simple ratio likely to be of interest to structurists (See also *ADDENDA*, § 3 −d)).

The break of the isotherm at 800°C that Picard and Dodé (23) obtained with *x* = 1.094 7 occurs almost exactly close to the boundary W'$_2$/W'$_3$. Finally, the limits *x*$_o$ and *x*$_1$ given by Campserveux *et alii* (37) and Touzelin (16) are often close to those found by Raccah.

## IV. – SUMMARY - CONCLUSION

In this memory is presented essentially the equilibrium phase diagram of the solid wüstite FeO$_x$ (see p. 18 Key versions 1964, 1979 and 2018) explaining how it was derived principally from the results of Raccah's thermogravimetric measurements using the phenomenological thermodynamics.

A general description of this 'classical' diagram is given on both sides of Θ = 911 °C while citing the experimental works that verified it. This memory will have been completed further (1986) by giving the tabulation of the molar thermodynamic properties of the solid wüstites, which will have been deducted principally from the same experimental results (See

*ADDENDA*, § 3 −d)). It is completed too by several references from the literature, some of them which accepted (or not) the pseudo phases and confirmed (or not) experimentally their existence, including a reviewing paper (2018) by the present authors.

So the present phase diagram constitutes a framework for the study of microscopic crystal and electronic properties (See ENDNOTES: contributions such as those on conductivity before and after quenching) and those of [Smyth, Goodenough, Collongues, Men', Toft Sørensen, Mrowec, Cohen, Lykasov, Stokłosa, … and others]' groups.

From the end of the 50s, the development of the chemical thermodynamics of defects concerning free energy and non-stoichiometry extended at the same time as that of the crystal structure determination. Point defects and clusters will have been taken into consideration as elements of the crystal lattice. Specific *diagrams of the defect structure* will have appeared progressively.

––––––––––––––––

## V. – *ADDENDA*

**1** − Metastability. Sorting the isotherms for identifying varieties.

Figures a) and b) below give a graphical representation of the isotherms $l'(x)$ determined by Raccah (Thesis 1962). The identification of the $W_i$ in fig. a) was simply made when considering the display of $M(T^{-1})$ and $N(T^{-1})$ in Figs. 6 and 7 p. 12 and 13 of paper (1). It was obvious as soon as 1962.

It was much more different for the $W'_i$ in Fig. b) for which it was carried out according to a large number of tests and errors, and validations or invalidations by least squares adjustments of the relation $l'(x_i)$ for each segment.

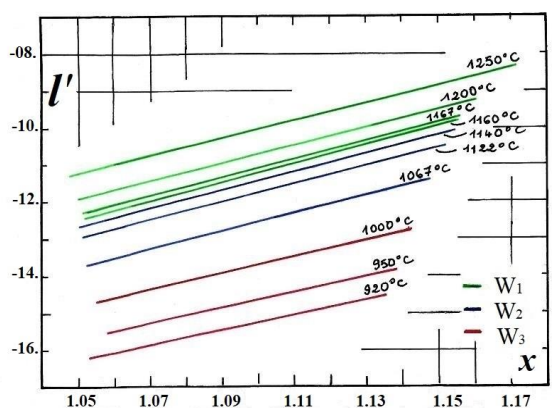 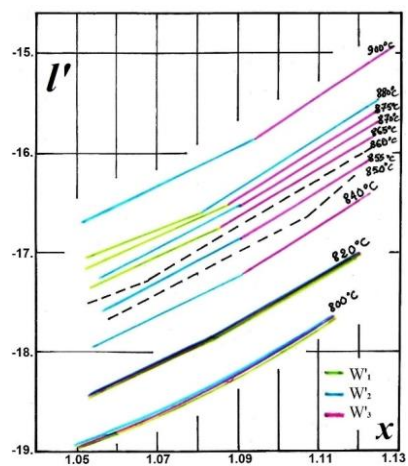

**a) Isotherms $l'(x)$ of W, T ∈[920-1250] °C**  **b) Isotherms $l'(x)$ of W', T ∈[800-900] °C**

The independent temperature value of the coefficients of each of the varieties $W'_i$ was established in this way despite the metastability of some segments not foreseeable in a specific temperature and composition range.

**2** − Key versions (1964) and (2018) of the equilibrium diagram via the present version (1979).

The first version (Fig. **A**) known as the 'Gallic Cocq' was published in CRAS Paris (1964) by Vallet – Raccah (3) p. 367, in French.

The last version (Fig. Ω) will have been published in English in 2018-19 (see Gavarri-Carel hal-01933760,v1 p. 8):

**Historical VERSION A (1964) → PRESENT VERSION 1979 → VERSION Ω (2018)**

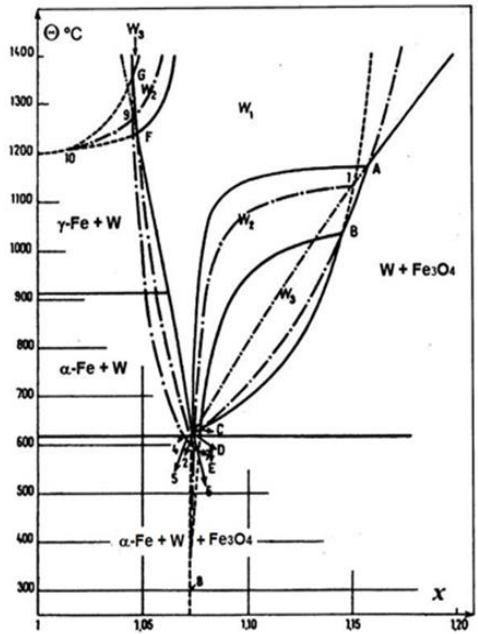

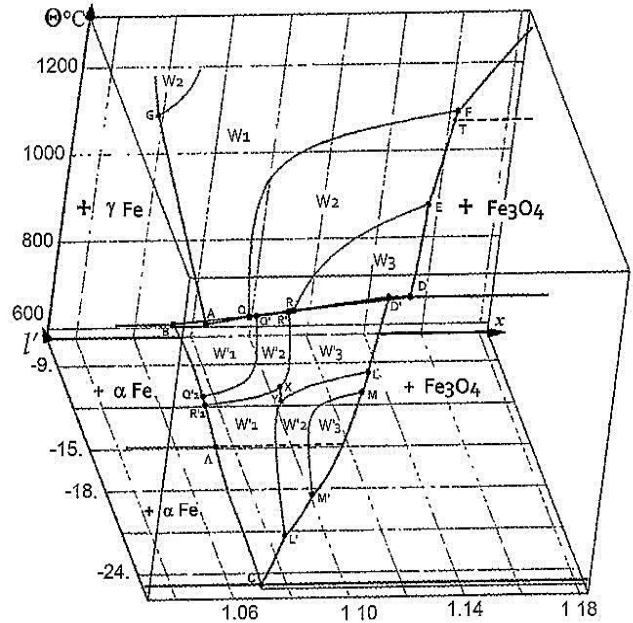

**A – Phase diagram 2D - $\Theta(x)$**
« LE COQ GAULOIS » (3) CRAS Paris (1964)

**Ω – Phase diagram 3D - $\Theta(l', x)$**
hal-01933760 Fig. 4 p. 8 (2018)

A short version allowing to draw roughly the phase diagram in the present version is available:
P. *Vallet, C. Carel, The Fe-O phase diagram in the range of the non-stoichiometric monoxide and magnetite at the Fe-rich limit. J. of Phase Equilibria (1989)* //doi.org/10.1007/BF02877494

**3 –** Some complements *post* 1979. About singularities and thermodynamic anomalies.

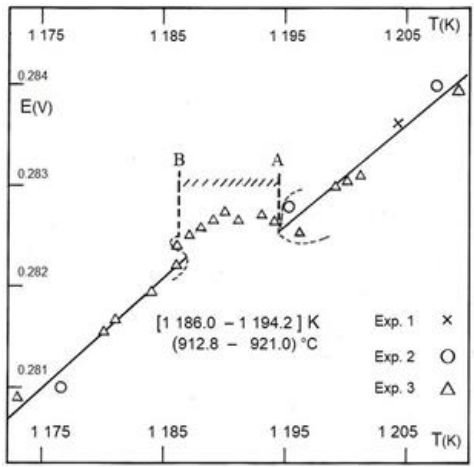

a) Points Q, Q', S', R', R, (D', D) of the isotherm at 1 184 K in Fig. 4. **The transition $W'_1 \leftrightarrows W_1$.**

This isotherm not only cannot be double, but the temperature of these points is probably not the same for all of them, nor precisely that of the $\alpha \leftrightarrows \gamma$ iron transition, except to be close to that of points A and B.
The first order transformation $W'_1 \leftrightarrows W_1$ will have been questioned by Wriedt in 1991 (*cf. infra* Ref.). In 1986, an exceptionally numerous series of very accurate values of the emf of a solid electrolyte battery along the Fe/W boundary will have been provided by Matts Hillert's group.

An anharmonic zone that appears –when plotting their data *vs* T in the temperature range [1 186.0 − 1 194.2] K or [912.8 − 921.0] °C will then allow to model an interpretation of the missing part of the diagram of the two wüstites –W'$_1$ -point B and W$_1$ -point A– which form a diphasic domain with α-Fe (+ε′$_O$) and γ-Fe (+ε$_O$) respectively at temperatures between 913 °C and 921 °C (*cf. infra* Ref. J.-R. Gavarri and C. Carel).

Ref. : -H. A. Wriedt (1991) The Fe-O (Iron-Oxygen) system  12 n°2 (1991) p. 176-177
doi.org/10.1007/10.1007/  BFO2645713

- O. Sjödén *et alii*, On the Gibbs energy of formation of wustite, Metall. Trans. B 17B (1986) p. 179-84, link.springer.com/article/10.1007_2FBF02670831-references.ris

- J.-R. Gavarri, C. Carel (2018). The complex  nonstoichiometry of  wüstite $Fe_{1-z}O$: Review and Comments, Progr.  Sol. St. Chem. **53** (2019) p. 27-49 (Fig. 3 p. 7)
doi.org/10.1016/j.progsolidstchem.2018.10.00 - free hal-01933760,v1

b) **The transition W′$_3$ ⇆ W$_3$** at ~1 184 K between D′ and D, and the $Fe_3O_4$ domain remains to be modelled, where the two wüstites W′$_3$ and W$_3$ with $Fe_3O_4$ would play a role similar to that of the wüstites W′$_1$ and W$_1$ with 'pure iron' (See $Fe_3O_4$ with multiple allotropic transformations, all probably not identified,  Ref. Wriedt (1991)).

- Carel, Vallet, Bull. Soc. Sci. Bretagne 52 (1-4) 1977-80, pub. 1981, then (2021), French; hal-03175135, English. More detailed in Vallet, Carel, Rev. Chim. miné. 24 (1987) p. 719, French. See reference in Lilova, Pear Ce, Gorski, Rosso, Navrotsky, Thermodynamics of the magnetite-ulvöspinel ($Fe_3O_4$-$Fe_2TiO_4$) Amer. Mineral. 97 (2012) p. 1330 *pdf downloadable*).

More simply (D,D′) could be the part (W′$_3$,W$_3$) of a triple point with $Fe_3O_4$.

c) **Triple points X and Y**. The equation of the common boundary W′$_1$/W′$_3$ in the coordinates (T, $x$) is as following

$$T = \frac{298\ 646.901\ x^2 - 499\ 256.362\ x + 190\ 009.684}{259.166\ 154\ x^2 - 424.588\ 558\ x + 155.487\ 327}$$

The curves admit a vertical asymptote for $x = 1.085\ 68$ which cancels the denominator. But for this value the numerator takes a low value in relation to the numbers that form it, *i.e.* -6.95, the canceling value of $x$ being 1.085 73. In other words, for either of these two values of $x$ or for their arithmetic mean $x = 1.085\ 7$, T takes the form 0/0, that is to say that any value of T is possible for this value of $x$, which means that the relation $l′(W′_1) = l′(W′_3)$ is checked along this asymptote. This relationship corresponds to the transformation of the second order W′$_1$⇆W′$_3$, so to the boundary W′$_1$/ W′$_3$. It results in the two invariant points X and Y (see Table 5 and Fig. 5).

d) **Thermodynamic properties** of the wüstites W$_i$ and W′$_i$ from Thermogravimetric Data at Equilibrium, P. Vallet, C. Carel: H$_i$, S$_i$, C$_{P,i}$; see new rectification of the α-Fe/W′ boundary from which it results a better localization of the **Chaudron's point C.** English.    hal-03083695 (2021)

**Point T** in Figs. 1 and 2: it is described in Ref. : C. Carel, P. Vallet, Bull. Soc. Sci. Bretagne 52 1977-1980, hal-03175135 (2021).

See also in the extended version *« P. Vallet, C. Carel, Propriétés thermodynamiques molaires et transformations dans la magnétite non stoechiométrique en équilibre avec les wüstites, Rev(ue) de) Chim(ie) miné(rale) 24 n° 6 (1987) p. 719-37 ; thermodynamic properties: 46 equations, Tables I and II. Ref. 0035-1032/87/6 719 19 Gauthier-Villars".* French.

Some elements of the diagram of $Fe_{3-δ}O_4$ in the Fig. 1 are from R. Dieckmann, Ber. Bunsenges. Phys. Chem. *86* (1982) 112-8 (Figs. 7 p. 118 in coordinates $l′$, $T^{-1}$).

See also in the paper hal-03175135, ENDNOTE–Additional references p. 4: the boundary W/$Fe_3O_4$, its analyze in the vicinity of 1 175 °C, and H. A. Wriedt's comments. Pending question: are points F and T distinct?

| Point | phases | T (K) | Θ °C | $x$ | $l'$ |
|---|---|---|---|---|---|
| B | α-Fe, γ-Fe, $W'_1$ | 1 184 | 911 | 1.049 5 | -16.426 |
| $Q'_1$ | α-Fe, $W'_1$, $W'_2$ | 1 184 | 911 | 1.058 3 | -18.571 |
| S | α-Fe, $W'_1$, $W'_3$ | 1 078 | 805 | 1.058 5 | -18.702 |
| $R'_1$ | α-Fe, $W'_2$, $W'_1$ | 1 074 | 801 | 1.058 9 | -18.809 |
| **C** | **$W'_1$, α-Fe, $Fe_3O_4$** | **864.73** | **591.6** | **1.070 0$_5$** | **-25.125** |

in *Rev. Chim. miné.* (1986). French => [hal-03083695: eq. (80): $l'_o$, with inflexion point at 748 °C, Tables VII and VIII p. 22 and 23 respectively (2021). English].

# VI. − ENDNOTES

**1** − Reception of the "varieties"

> '*Der Verlauf des Sauerstoffpartialdruckes als funktion der Zusammensetzung für FeO kann nicht mehr mit Hilfe eines einfachen Fehlordnungsmodells gedeutet werden*'
> Von *H.-G. SOCKEL und H. SCHMALZRIED*

[Coulometrishe Titration an Übergangsmetalloxiden, Zeit. Phys. Chem. N. F. Bd. 72 Nr. 7 (1968) S. 745. English]

The affirmation by Wagner jr *et alii*, and Fender *et alii*, in 1966 and 1969 respectively, when identifying the pseudo phases, and the identity of their results with those by Raccah-Vallet-Carel (1962-65) contradicts punctually the famous Max Planck's remark: see Vallet, Carel (2021): hal-03083695 'Epistemological truth' Endnote 3 - p. 25.

The pseudo phases and their properties will have taken place by conductivity studies:
− In Soviet Union, S. M. Ariya and B. Ya. Bratch, Electrical Conductivity of Ferrous Oxide at High Temperatures (600 − 900 °C), Fiz. Tverd. Tela $\underline{5}$ N°2 (1963) 3499. Russian. Translation: Soviet Physics–Solid State, $\underline{5}$ N° 12 (1964) p. 2565. See measurements *after quenching* p. 2566 Fig. 1 in coordinates ($x_{Fe3+}$ (~$2z$), conductivity σ).

 − V. A. Kozheurov, G. G. Mikhailov, Electrical Conductivity of Wüstite, Russian Journal of Physical Chemistry (English translation) $\underline{41}$ N° 11 (1967) p. 1552. *After quenching* in Fig. 1 p. 1554 σ = f($x_{Fe3+}$)

− In U.S.A., Wagner *et alii* ( (1966, Ref. (10)) Electrical conductivity of undoped wüstite, *under equilibrium*: p. 950 Fig. 1 in coordinates (Log σ, 2 Log ($P_{CO2}/P_{CO}$)); despite the crushing due to logarithms and spreading of the $x$-axis, fluctuations can be observed as undulations corresponding to the transitions $W_2/W_3$ but not $W_1/W_2$. That is a strange interpretation of their results in *Table* 4 p. 955 (!)

New defect thermodynamics was explored and developed:

− In U.K. D. M. Smyth, Deviations from stoichiometry in MnO and FeO, J. Phys. Chem. Solids $\underline{19}$ (1/2) (1961) p. 167. The proportionality between the stoichiometric deviation and $p_{O2}^{1/n}$, p. 168 $MnO_{1+x}$: the concentration (the molar rate) of cation vacancies [□] = v is equal to $x/(1+x)$ ∝ $p_{O_2}^{1/6}$. See also  D. M. Smyth, The defect chemistry of metal oxide, Chapter 13  NiO and CoO  p. 239-52, Oxford Univ. Press (2000).

− In France, M. Dodé *et alii*, (1970) Ref. (25)

− J. B. Goodenough, (1971), Ref. (43)

− As soon as 1972 *in* UK N. N. Greenwood, A. T. Howe, Mössbauer studies of $Fe_{1-x}O$. 1. The defect structure of quenched samples; 2. Disproportionation between 300 and 700 K; 3. Diffusion line broadening at 1 074 and 1 173 K, J. Chem. Soc. Dalton Trans.1 (1972) p. 110, 116, 122-26 doi.org/10.1039/DT9720000110, doi.org/10.1039/DT9720000116, doi.org/10.1039/DT9720000122

− W. Burgmann, The defect structure models for Wüstite: a review, Met. Sci. (41)
doi.org/10.1179/030634575790445107

− In France E. Bauer, A. Pianelli, Mat. Res. Bull. + A. Aubry, F. Jeannot, A review of the lacunar structure of wüstite, and the necessity of new experimental researches + New structural examination of pure and substituted metastable wüstites, Mater. Res. Bull. 15 n° 2 (1980) p. 177-88, and n° 3 p. 323-37.

− In Japan E. Takayama, N. Kimizuka, Thermodynamic properties and subphases of wustite field determined by means of thermogravimetric method in the temperature range 1 100-1 300 °C, J. Electro-chem. Soc. 127 (4) (1980) p. 970.
https://scholar.google.fr/scholar?hl=fr&as_sdt=0%2C5&q=Takayama +E%2C+Kimizuka+N&btn

− In the Danish O. Toft Sørensen's group, Nonstoichiometric Oxides, O. Toft Sørensen Ed., Acad. Press, New York (1981), the wüstite p. 39-44. With El Sayed Ali, Defects on metal-deficient oxides: wüstite, $Fe_{1-y}O$, Rep. 4000 Roskilde Risø-R-505 The Risø National Lab. (Jan. 1985) p.1-24
ISSN 0106-2840

 − At Washington and Berkeley (USA), respectively R. M. Hazen and R. Jeanloz, Rev. geophys. and space Phys. $Fe_{1-x}O$: a review of its defect structure and physical structure and physical properties 22 (1984) p. 22-37    //doi.org/10.1029/RG022i001p00037

− C. Gleiser in collaboration with J. B. Goodenough, Mixed-valence iron oxides in Structure and bonding Vol. 61 Springer Verlag (1985) p. 51-4. C. Gleiser, Key Eng. Mater. 125-6 (1997) p. 359-380

− In the Swedish thermodynamician Mats Hillert's Group, a very accurate work about the first order transformation W⇄W' on the boundary ($\alpha$ or $\gamma$Fe)/(W' or W): O. Sjödén, S. Seetharaman, L.-I. Staffansson (See above § 3, a))         //doi.org/10.1007/BF02670831]

− At the Mining and Metallurgy Academy (AGH) Cracow-Poland:
− In J. Janowski, St. Jasienska' group: Ref. (29); J. Janowski, S. Mrowec, A. Stokłosa, Determination of Chemical Diffusion and Self-diffusion Coefficients of Iron in Ferrous Oxide, Bulletin de l'Academie Polonaise des Sciences, Série des sciences chimiques XXIX N°1-2 (1981) Pub 1982    p. 91 ; J.-R. Gavarri, St. Jasienska, J. Orewczyk, Contribution to the study of substituted wüstites $Fe_{1-x-y}O$, Metalurgia I Odlewnictwo Krakow-Poland 13 (1-2) 1987, p. 43

− In S. Mrowec's group at AGH, with A. Podgorecka, Defect structure and transport properties of non-stoichiometric ferrous oxide, Review. J. Mater. Sci. 22 (1987) p. 4181.

− A. Stokłosa (AGH-Cracow) in J. J. Molenda, A. Stokłosa, W. Znamirowski, Physica Status Solidi (b) 142 (1987) p. 517. The three pseudophases w1, w2, w3 are identified (Fig. 4 p. 522) in the graph of conductivity (measurement under equilibrium) *vs* departure of stoichiometry. See particularly the isotherm at 1274 K. A significant remark about the possible nature of a transition of type semiconductor-metal W1⇆ (W2-W3) is due to J. Molenda.
 //doi.org/10.1002/pssb.2221420221

− Men's Group (Institute of Metallurgy at Sverdlovsk (now Ekatrinburg–Russia), and Lykasov at Tchelyabinsk − Russia as referenced specialists of the thermodynamics of metal solutions in oxides: they will have taken in hand the writing of a book: A. A. Lykasov, C. Carel, A. N. Men', M. T. Varshavskii, G. G. Mikhailov, Fiziko-Chimitcheskie Svoistva Viustita I Ego Rastvorov (Physicochemical Properties of Wüstite and its Solutions), Instituta Metallurgii I Riso Yunts Akademiia Nauka SSSR, Sverdlovsk GSP-169 (1987) 230 pp.

− in France, R. Collongues, Nonstoichiometry in oxides, about the wüstite, Prog. Cryst. Growth Char. 25 (1992) p. 209-14

 − … and some other authors.

**2** − *In 2010, the Canadian metallurgists*: E. J. Worral and K. S. Coley published *the third indisputable experimental verification* of the 3x2 varieties $W_i$ and $W'_i$ at 850 and 1050 °C, by means of a kinetical method concerning the carbon isotopes $^{12}C$ and $^{13}C$. The process of this method was reactivated because it has become technically very sensitive. These authors referred systematically to the discover of the $W_i$ and $W'_i$ which they identified again: E. J. Worral, K. S. Coley, 1. Kinetics of the reaction of $CO_2/CO$ gas mixtures with iron oxide, Metall. Mater. Trans. B, <u>41B (2010)</u> p. 813-23       //doi.org/10.1007/s11663-010-9358-4
Remark: their isotherm at 850 °C evidences three $W'_i$, not only two as it was sayed following first Wagner, jr's affirmation (see hal-01933760,v1, Annex B  2-"Re-analysis of kinetic data by Worral and Coley" p. 36)
       2. Defect structure of pseudo-phases of wüstite, Canadian Metallurgical Quarterly 52 N°1 (2013) p. 23-33          //doi 10.1179/1879139512Y.0000000047

**3** − *A significant improved review of the pseudo phases defect structure*: A. Stokłosa (Cracow-Poland), Non-Stoichiometric Oxides of 3D-Metals (Diagrams of the Concentration of Point Defects), *Vol 79 (2015) of Mat. Sc. Found. Iron oxide Fe1-$\delta$ O p. 313-75, Trans. Tech. Pub. Ltd-Switzerland*,                  //doi: *10.4028/www.scientific.net/MSFo.79.313*:
A powerful thermodynamic description of the defect structure (point and complex defects) of the pseudo phases with their electronic properties is deducted from the relation between the rate of non-stoichiometry and the oxygen pressure at equilibrium: $\delta$ (≡**z**) = $p(O_2)^{1/n}$. The Ref. of this work is missing in the bibliography of J.-R. Gavarri, C. Carel (2018), The complex nonstoichiometry of wüstite Fe1-**z**O: Review and comments, Prog. in Sol. St. Chem. <u>53</u> (2019) p. 27 (free hal-01933760,v1). It should be added in a next version.

**4** − Nevertheless, the wüstite was studied before 1979 (and after) in some big labs.

It was the case at the Northwestern University (Evanston near Chicago - U.S.A.) from 1967, where Cohen's group equipped with advanced experimental devices will have continued until 1993 the search about the crystalline structure of clusters of point defects without identifying some varieties. For the first time in 1969, Koch and Cohen (42) described the cluster (13/4) whose $3a_O$ repetition factor allows a refinement of a commensurate crystal structure of a wüstite only with composition given by $x$ = 0.098.
In the course of a study in T. O. Mason's group at the Northwestern too, taking back an early work −W. J. Hillegas's thesis, 1968− E. Gartstein worked on electrical properties (1962, conductivity and Seebeck effect). In two papers, the main part of the research focused on the most likely cluster following the value of $x$ in Fe$_{1-x}$O. At the end, they will have introduced the cluster (5/2)* appearing at low $x$, the historical (13/4) cluster remaining the major one at large $x$.

Refs. E. Gartstein, T. O. Mason, J. B. Cohen, Defect agglomeration in wüstite at high temperatures −I. The defect arrangement. E. Gartstein, J. B. Cohen, T. O. Mason, −II. An electrical conduction model, J. Phys. Chem. Solids 47 N° 8 (1986) p. 759-773 and 775-781.

*(m/n) = 2.5: Citing S. Iijima, Elect. Micros. Soc. Amer. 32 (1974) p. 619-23, and A. B. Anderson, R. W. Grimes, A. H. Heuer, J. Sol. St. Chem. 55 (1984) p. 353-61, J.-R. Gavarri (CRAS-Paris, tome 306 Série II (1988) p. 957-962) will have introduced the relation (m-n) = $4zk^3$. After identification in the literature, he will have justified the cluster* (10/4) blende ZnS as the most convenient elemental complex structure with an experimental repetition factor 2.4 whatever is the composition.

Opposite views,
- Beyond 1966, the thermodynamician Wagner jr (in Cohen's Group), and 1969 members of Fender's group will no longer report modifications in their numerous subsequent publications concerning the wüstite. The first authors focused then on crystal structure and surface properties without transitions, the later ones on theoretical calculations of energy of formation in the clusters. The varieties were never more envisaged. In both cases, their attitude will be become in agreement with the Max Planck's assertion.
- In Cohen's group, it will have been maintained until 1993 a crystallographic search about the wüstite and the manganosite, avoiding to take pseudo phases into consideration. So it was in constant agreement with the Max Planck's assertion.
- In France, in Dodé's Group, G. Gerdanian *et alii* didn't meet the "varieties" when measuring with a calorimeter the ratio $\delta q/\delta n_{02}$ ($\cong \epsilon/\epsilon'$) for a $\delta n_{02}$ value as most minimized as possible, nevertheless considering the value of $\overline{H}_O - \frac{1}{2}H°_{O2}$ obtained so as more accurately determined as that obtained with derivation of the equilibrium value $\overline{G}_O - 1/2G°_{O2}$ known by measurement of the oxygen potentiel (log $p_{O2}$), and the composition $z$ of the Fe$_{1-z}$O/O$_2$ equilibrium (P. Gerdanian, Thermodynamic study of non-stoichiometric oxides by high temperarure microcalorimetry, Advances in Solid-State Chemistry 1 (1989) p. 225-58 ISBN: 0-89232-867-3)
- In Cracow-Poland, J. Nowotny, St. Bialas, I. Sikora, Metalurgia I Odlewnietwo 6 n°2 (1980) p. 161-170, published a paper based in mixed manner on recognition of validity of some identifications of the pseudo phases by Vallet's group, and on statistical calculations invalidating their existence. Nowotny's work will have continued without meeting any pseudo phase until 2020 concerning the surface properties of W' only (for examples with J. B. Wagner jr in Oxyde Metals, 15 1/2 (1981) p. 169, and with T. Bak by means of the tool of 'work functions' (*ACS Appl. Energy Mater.* 2020, 3, 10, 9809).

**Aknowledgements.** Our thanks to G. Urbain of the « **I**(nstitut de) **R**(echerche) (de la) **Sid**(érurgie) » at Saint-Germain-en-Laye-France, for having allowed the reproduction in the chemistry department at the University of Rennes of a 'gas line' for the purification and the flow measurement of gases CO and CO2. This wüstite study will be indebted to the staff of the Institute Laue-Langevin (Grenoble-France) including its Director B. Jacrot and the Chief-engineer G. Gobert, who facilitated the conception, construction and funding of a 'diffractofurnace', so the continuation of the defect structure study under equilibrium by neutron diffraction. Many thanks to Professor D. Weigel for his continuous participation to our research.

———————————————

# REFERENCES

(∅) P. Vallet, C. Carel, Contribution à l'étude du protoxyde de fer solide non stoe-chiométrique. Diagramme T-P-X, Mat. Res. Bull. <u>14</u> N° 9 (1979) p. 1181-1194
//doi.org/10.1016/0025-5408(79)90213-7

(1) P. Vallet, P. Raccah, Rev. Métall., Mém. Sci. <u>62</u> (1965) p. 1

(2) P. Raccah, Thèse série B n° 7 n° d'ordre 8, Rennes-France (1962)

(3) P. Vallet, P. Raccah, Compt. Rend. Acad. Sci. France <u>258</u> (1964) p. 367
http://gallica.bnf.fr/ark:/12148/12148/bpt6k4011c.image.r=comptes+rendus+academie+sciences+paris+1963.f1301.langFR.pagination

(4) P. Raccah, P. Vallet, Compt. Rend. Acad. Sci. Fr. <u>253</u> (1961) p. 2682, and <u>254</u> (1962)
        p. 1038. With M. Kléman, Compt. Rend. Acad. Sci. Fr. <u>256</u> (1963) p. 136

(5) C. Carel, Thèse série B n° 27 n° d'ordre 58, Rennes-France (1966). Rev. Métall. Mém.
        Sci. <u>64</u>, n° 9 p. 737 and n° 10 p. 821 (1967)

(6) J.-R. Gavarri, Thèse série B, Paris 6 (25 septembre 1978)

(7) C. Carel, J.-R. Gavarri, Mat. Res. Bull. <u>11</u> (1976) p. 745. J.-R. Gavarri, D. Weigel, C.  Carel,
        Mat. Res. Bull. <u>11</u> (1976) p. 917

(8) J.-R. Gavarri, C. Berthet, C. Carel, D. Weigel, Compt. Rend. Acad. Sci. Fr. <u>C285</u> (1977)
        p. 237. J.-R. Gavarri, C. Carel,  D. Weigel, J. Solid. St. Chem. 28 n° 5 (1979) p. 81.
        Short range order in solid wüstite, Comm. Cong. Appl. Cryst., Porabka-Kozubnik
        Aug. 1978, Silesian University of Katowice, Inst. Ferrous Metall. Gliwice – Pologne,
        Proceed. II (1979) p. 832

(9) C. Carel, P. Vallet, Compt. Rend. Acad. Sci. Paris  a) <u>258</u> (1964) p. 3281, b) J.-P. Bars, C.
        Carel, *ibidem* <u>C269</u> (1969) p. 1152, c) C. Carel, *ibidem* <u>273</u> (1971) p. 393

(10) G. H. Geiger, R. L. Levin, J. B. Wagner jr, J. Phys. Chem. Solid  27 (1966) p. 947.
        //doi.org/10.1016/0022-3697(66)90066-7

(11) B. E. F. Fender, F. D. Riley, J. Phys. Chem. Solids <u>50</u> (1969) p. 793
        //doi.org/10.1016/0022-3697(69)90273-X

(12)  M. Hayakawa, Ph. D. Thesis, X-ray Diffraction Studies of Wüstite at High Temperature,
        Northwestern Univ. Evanston Illinois (1973). Same author, J. B. Cohen, T. B. Reed,
        J. Amer. Ceram. Soc. <u>55</u> (1972) p. 160. Same author, J. B. Wagner jr, J. B. Cohen,
        Mat. Res. Bull. <u>12</u> (1977) p. 429

(13) J. Manenc, J. Phys. et le Rad. <u>24</u> (1963) p. 447. T. Herai, J. Manenc, Mém. Sci. Rev.
        Métall. <u>LXI</u> (1964) p. 677 . Avec B. Thomas, *ibidem* <u>LXIII</u> (1966) p. 397. J.
        Manenc, Bull. Soc. Fr. Mineralog. Cristallogr. <u>97</u> (1968) p. 594

(14) B. Andersson, J. O. Sletnes, Acta Cryst. <u>A33</u> (1977) p. 268

(15) P. Vallet, Compt. Rend. Acad. Sci. Fr. <u>261</u> (1965) p. 4396

 (16) B. Touzelin, *Etude par la diffraction des rayons X à haute température et en atmosphère
        controlée des oxydes non stoechiométriques FeO$_{1+x}$ et MnO$_{1+x}$*, Thèse série A, n°
        d'ordre 1302, Orsay 1974

(17) H. F. Rizzo, Thermodynamics of the Fe-O system by Coulometric Titration in High
        Temperature Galvanic Cells, Ph. D. Thesis, University of Utah (1968)

(18) P. J. Spencer, O. Kubaschewski, CALPHAD 2, Pergamon Press  2 (1978) p. 147

(19) T. I. Boulgakova, O. S. Zaytsev, A. G. Rozanov, Vestn. Moskov. Univ. <u>3</u> (1966) p. 102

(20) B. Swaroop, J. B. Wagner jr, Trans. Met. Soc. AIME <u>239</u> (1967) p. 1215

(21) I. Bransky, A. Z. Hed, J. Amer. Ceram. Soc., Discuss. and Notes <u>51</u> (1968) p. 231

(22) H. Asao, K. Ono, A. Yamaguchi, J. Moriyama, Mem. Fac. Engineer., Kyoto Univ. $\underline{32}$ part 1 (1970) p. 66

(23) C. Picard, M. Dodé, Bull. Soc. Chim. Fr. (1970) p. 2486

(24) J.-Y. Boudonnet, Thèse série B n° 52, n° d'ordre 73 Rennes-Fr (1968). Avec C. Carel, Mém. Sci. Rev. Metallurg. LXX N° 3 (1973) p. 179

(25) R. A. Giddings, R. S. Gordon, J. Amer. Cer. Soc. $\underline{56}$ (1973) p. 111

(26) K. Löberg, W. Stannek, Ber. Buns. Ges. Phys. Chem. $\underline{79}$ (1975) p. 244

(27) J. Janowski, R. Benesh, M. Jaworski, A. Miklasinski, Pr. Kom. Ceram. Pol. Akad. Nauk. $\underline{21}$ (1974) p. 139. With R. Kopec, A. Wilkosz, Kom. Metal. Odlev, $\underline{22}$ (1974) p. 65

(28) B. Leroy, G. Béranger, P. Lacombe, J. Phys. Chem. Solids $\underline{33}$ (1972) p. 1515

(29) P. Vallet, C. Carel, Ann. Chim. $\underline{5}$ (1970) p. 246

(30) P. Vallet, Compt. Rend. Acad. Sci. Fr. a) $\underline{C280}$ (1975) p. 239, b) $\underline{C281}$ (1975) p. 291, c) $\underline{C284}$ (1977) p. 545

(31) L. S. Darken, R. W. Gurry, J. Amer. Chem. Soc. $\underline{67}$ (1945) p. 1398.

(32) J. Nowotny, I. Sikora, Bull. Acad. Sc. Pol. $\underline{23}$, 12 (1975) p. 1045, Zeit. Physik. Chem. Neue Folge $\underline{107}$ (1977) p. 587, and J. Electrochem. Soc. $\underline{125}$ 5 (1978) p. 781

(33) G. Chaudron, Thèse série A n° 875, n° d'ordre 1689, Paris (1921)

(34) J. Bonneté, J. Païdassi, communication privée.

(35) F. D. Rossini, Fundamental Measures and Constants for Sciences and Technology, C.R.C. Press Cleveland-Ohio, 100 (1974)

(36) C. Carel, D. Weigel, P. Vallet, Compt. Rend. Acad. Sci. Fr. $\underline{258}$ (1964) p. 6126, and $\underline{260}$ (1965) 4325

(37) J. Campserveux, G. Boureau, C. Picard, P. Gerdanian, Rev. Internat. H. Temp. Réfract. $\underline{6}$ (1969) p. 185

(38) D. M. Smyth, Deviation from stoichiometry in MnO and FeO, J. Phys. Chem. Solids $\underline{19}$ (1/2) (1961) p. 167; [cation vacancies rate $\propto p_{O_2}^{1/n}$, p. 168 $MnO_{1+x}$ : [$\square$] $\propto p_{O_2}^{1/6}$]. The defect chemistry of metal oxides, N. Y. Oxford Univ. Press (2000) Book, pp. 299

(39) R. Collongues, Thèse série A n° de série 2934 n° d'ordre 5805, Paris 1954, and La Non-Stoechiométrie, Monographie de Chimie, Masson, Paris 23 (1971). Nonstoichiomtry in oxides. Prog. Cryst. Growth. Char. $\underline{25}$ (1992) p. 203, about the wüstite 209-214.

(40) J. B. Goodenough, Metallic Oxides, Pergamon Press (1971). Enhanced Edition, translated into French by A. Casalot « Les Oxydes de Métaux de Transition », Gauthier-Villars, Paris (1973) p. 166. C. Gleiser, J. B. Goodenough, Mixed-valence iron oxides in structure and bonding vol. 61 (1985) Springer Verlag p. 51-4

(41) W. Burgmann, The defect structure models for Wüstite: a review, Met. Sci. 9 (1975) p. 169
doi.org/10.1179/030634575790445107

(42) F. Koch, J. B. Cohen, Acta Cryst. $\underline{B25}$ (1969) p. 275. M. Hayakawa, M. Morinaga, J.B. Cohen, Defects and Transport in Oxides, Eds M. S. Seltzer and R. I. Jaffee, Plenum Press (1974)

————————————————